# EXTRACTION OF HADRON-HADRON POTENTIALS ON THE LATTICE WITHIN 2+1 DIMENSIONAL QED [1]

H.R. FIEBIG [a], O. LINSUAIN [a], H. MARKUM [b], K. RABITSCH [b]

[a] *Physics Department, F.I.U. - University Park*
*Miami, Florida 33199, U.S.A.*
[b] *Institut für Kernphysik, Technische Universität Wien*
*A-1040 Vienna, Austria*

A potential between mesons is extracted from 4-point functions within lattice gauge theory taking 2+1 dimensional QED as an example. This theory possesses confinement and dynamical fermions. The resulting meson-meson potential has a short-ranged hard repulsive core due to antisymmetrization. The expected dipole-dipole forces lead to attraction at intermediate distances. Sea quarks lead to a softer form of the total potential.

---

[1]Supported in part by the National Science Foundation Contract PHY-9409195, by the "Fonds zur Förderung der wissenschaftlichen Forschung" Contract P10468-PHY, and by CEBAF.

*1. Introduction.* The extraction of an effective interaction, or potential, between two composite hadrons from a quantum field theory is of fundamental importance. On one hand, since a potential is a central object of classical and quantum mechanics, its definition based on a quantum field is interesting in itself. On the other hand, the potential between clusters of particles has proved very useful in many studies of subatomic physics. Here we have in mind the quark-gluon degrees of freedom at the subnuclear level which are thought to give rise to the nucleon-nucleon forces usually described by meson theory. The vacuum of QCD contains both virtual gluons and quarks. At short distances the nucleon-nucleon forces are mediated by gluon exchange between the constituent quarks whereas for longer distances the production of quark-antiquark pairs should be the dominating mechanism. Quark-antiquark exchange can be treated as an effective meson exchange leading to the construction of the Bonn and Paris potentials [1]. Gluon exchange can be studied by phenomenological potential and bag models which provide some insight into the interaction mechanism of the six-quark system [2].

The aim should be to calculate the nucleon-nucleon forces directly from the quantum fields of QCD. Such a project has to deal with the *effective forces between composite colour singlets*, in other words, their "chemistry". This goes along with a characteristic *change of scale* leaving a residual effective interaction $10^{-2}$ to $10^{-3}$ times smaller than a typical hadron mass. The numerical difficulties involved in going from the GeV to the MeV scale in a lattice computation are evident. For this reason it seems appropriate to investigate simpler lattice models which exhibit similar physics. Naturally, one needs to insist on confinement and the presence of fermions. In this spirit we report on a simulation aimed at extracting an effective meson-meson interaction from a QED gauge field model in $2+1$ dimensions with staggered fermions which has confinement.

In the last years lattice calculations with static quarks have demonstrated that the potential between two three-quark clusters yields an attractive potential limited to the overlap region of the baryons [3, 4]. A hard repulsive core of the potential, as stipulated by experiments and their interpretation, could not be observed in the region where the two nucleons have relative distance close to zero. Beyond the static theory, employing dynamical quark propagators, correct antisymmetrization and quark exchange become possible [5]. In this framework the mechanism which leads to a repulsive core will be investigated here analytically and numerically within lattice $QED_{2+1}$.

2. *Lattice QED.* We have studied a euclidean lattice field theory in $2 + 1$ space-time dimensions for a $U(1)$ gauge symmetry group and staggered fermions. The action of the system is $S(U, \bar{\chi}, \chi) = S_G(U) + S_F(U, \bar{\chi}, \chi)$ which depends on the gauge fields $U$ and on the fermion fields $\bar{\chi}, \chi$. For the gauge action $S_G$ we use Wilson's plaquette formulation

$$S_G = \beta \sum_{x,\mu<\nu} (1 - \frac{1}{N} \text{ Re Tr } U_\mu(x) U_\nu(x + \hat{\mu}) U_\mu^\dagger(x + \hat{\nu}) U_\nu^\dagger(x)) \ , \qquad (1)$$

where $\beta = 2N/g^2$ contains the gauge coupling $g$ and the number of colours $N$ which is one in our case. For the fermionic action $S_F$ we employ the Kogut-Susskind prescription [6]

$$S_F = \frac{N_F}{4} \left\{ \sum_{x\mu} \Gamma_{x\mu} \frac{1}{2} (\bar{\chi}(x) U_\mu(x) \chi(x + \hat{\mu}) - \bar{\chi}(x + \hat{\mu}) U_\mu^\dagger(x) \chi(x)) + m_F \sum_x \bar{\chi}(x) \chi(x) \right\} \ , \qquad (2)$$

where $N_F$ is the number of external flavours, $m_F$ the dynamical quark mass and the $\Gamma_{x\mu}$ are reminiscent of the Dirac matrices.

Expectation values of physical quantities, defined through path integrals, are estimated via Monte Carlo simulations. We compare results for $QED_{2+1}$ with and without dynamical fermions. The simulation was done on an $L^2 \times T = 24^2 \times 32$ periodic lattice with the compact $U(1)$ Wilson action at $\beta = 1.5$ and staggered fermions with $N_F = 0$ and $N_F = 2$, both with $m_F = 0.1$. We used 64 independent gauge field configurations generated with the molecular dynamics algorithm [7]. Fermion propagators were computed utilizing a random source technique [8] with 32 and 16 random sources in the quenched case and in full $QED_{2+1}$, respectively.

3. *Meson-Meson Correlator.* The one-meson field is a product of staggered Grass-

mann fields $\chi$ and $\bar{\chi}$ with external flavours $u$ and $d$

$$\phi_{\vec{x}}(t) = \bar{\chi}_d(\vec{x}\,t)\chi_u(\vec{x}\,t)\,. \tag{3}$$

The meson-meson fields with relative distance $\vec{r} = \vec{y} - \vec{x}$ are then constructed by

$$\Phi_{\vec{r}}(t) = L^{-2}\sum_{\vec{x}}\sum_{\vec{y}}\delta_{\vec{r},\vec{y}-\vec{x}}\phi_{\vec{x}}(t)\phi_{\vec{y}}(t) = L^{-2}\sum_{\vec{x}}\bar{\chi}_d(\vec{x}\,t)\chi_u(\vec{x}\,t)\bar{\chi}_d(\vec{r}+\vec{x}\,t)\chi_u(\vec{r}+\vec{x}\,t)\,. \tag{4}$$

The information about the dynamics of the meson-meson system is contained in the 4-point time correlation matrix

$$C^{(4)}_{\vec{r}\vec{r}'}(t,t_0) = \langle \Phi^\dagger_{\vec{r}}(t)\Phi_{\vec{r}'}(t_0)\rangle - \langle\Phi^\dagger_{\vec{r}}(t)\rangle\langle\Phi_{\vec{r}'}(t_0)\rangle\,, \tag{5}$$

where $\langle\ \rangle$ denotes the gauge field configuration average. Working out the contractions between the Grassmann fields the following diagrammatic representation is obtained

$$C^{(4)} = C^{(4A)} + C^{(4B)} - C^{(4C)} - C^{(4D)} \tag{6}$$

$$= \ \text{∥∥} \ + \ \text{⋈} \ - \ \text{⋈} \ - \ \text{⋈} \ . \tag{7}$$

Each of the four contributions to the correlator comprises the exchange of gluons and sea quarks. For diagrams $C^{(4A)}$ and $C^{(4B)}$ those take place between the mesons, whereas diagrams $C^{(4C)}$ and $C^{(4D)}$ correspond also to interaction mediated by the exchange of valence quarks. Denoting contractions of the Grassmann fields by

$$\ldots\overset{n}{\chi}\ldots\overset{n}{\bar{\chi}}\ldots = G_n\,, \quad \text{with} \quad n = 1\ldots 4\,, \tag{8}$$

we have for example

$$C^{(4A)} \sim \langle\overset{43}{\phi^\dagger_{\vec{y}'}}\,\overset{21}{\phi^\dagger_{\vec{x}'}},\overset{12}{\phi_{\vec{x}}}\,\overset{34}{\phi_{\vec{y}}}\rangle = \langle G_1 G_2^* G_3 G_4^*\rangle\,, \quad \text{with} \quad \vec{r} = \vec{y} - \vec{x}\,, \vec{r}' = \vec{y}' - \vec{x}'\,, \tag{9}$$

where $\sim$ stands for the sums with normalization factors that carry over from (4). The gauge configuration average is taken over the product of all four propagators $G$.

In order to define an effective meson-meson interaction it is crucial that the noninteracting components in $C^{(4)}$ are isolated. This is achieved by means of a cumulant (or cluster) expansion [9] of the gauge field average. For example

$$
\begin{aligned}
C^{(4A)} \quad \sim \quad & \langle G_1 G_2^* \rangle \langle G_3 G_4^* \rangle + \langle G_1 G_3 \rangle \langle G_2^* G_4^* \rangle \\
& + \langle G_1 G_4^* \rangle \langle G_3 G_2^* \rangle + \langle\!\langle G_1 G_2^* G_3 G_4^* \rangle\!\rangle \, ,
\end{aligned}
\tag{10}
$$

where the last term defines the cumulant. In the first term no interactions from gluons or sea quarks *between* the mesons exist. It is thus interpreted to describe freely propagating mesons on the lattice. With reference to (8)–(10) define

$$
\bar{C}^{(4A)} = \langle \phi_{\vec{x}}^{\dagger\,21}, \phi_{\vec{x}}^{\,12} \rangle \langle \phi_{\vec{y}}^{\dagger\,43}, \phi_{\vec{y}}^{\,34} \rangle \sim \langle G_1 G_2^* \rangle \langle G_3 G_4^* \rangle .
\tag{11}
$$

Diagram $C^{(4B)}$ can be analyzed in a similar fashion, which then leads us to define $\bar{C}^{(4B)}$. The sum of those

$$
\bar{C}^{(4)} = \bar{C}^{(4A)} + \bar{C}^{(4B)}
\tag{12}
$$

respects the boson symmetry on the meson-field level and constitutes the free meson-meson correlator with the *relative* interaction switched off.

*4. Effective Interaction from Time Evolution.* The deviation of $C^{(4)}$ from $\bar{C}^{(4)}$ contains the residual effective meson-meson interaction. To find its definition, one may calculate the correlation matrix $\hat{C}$ between two elementary boson fields $\hat{\phi}(\vec{x}t)$ that live on the lattice sites. The corresponding quantum mechanical Hamilton operator consists of a free and an interacting part, $\hat{H} = \hat{H}_0 + \hat{H}_I$. It is possible to derive an explicit expression for $\hat{H}_I$ in terms of the correlators of order $n = 0$ and $n = 1$ from the perturbative expansion of $\hat{C}$ fulfilling

$$
\hat{C}(t - t_0) = e^{-\hat{H}_I (t - t_0)} .
\tag{13}
$$

This result leads us, by way of analogy, to define an effective meson-meson interaction as

$$
\mathcal{H}_I = - \frac{\partial \mathcal{C}}{\partial t} \bigg|_{t = t_0} , \quad \text{with} \quad \mathcal{C} = \bar{C}^{(4) - \frac{1}{2}} C^{(4)} \bar{C}^{(4) - \frac{1}{2}} .
\tag{14}
$$

This definition is meant as a relation between matrices and holds independently of the basis. It can be tested both in momentum space [10] and in coordinate space. The above equations build a bridge between the quantum field theoretical correlation functions on the lattice and an effective quantum mechanical Hamiltonian.

5. *Hard Core from Adiabatic Approximation.* An analysis of the coordinate-space matrix elements of the correlators $C^{(4)}$ and $\bar{C}^{(4)}$ reveals an interesting perspective on the repulsive core. From $C^{(4)}$ expressed in terms of the fermion propagators it is easy to see *analytically* that

$$C^{(4)}_{\vec{r}\vec{r}'}(t, t_0) = C^{(4A)}_{\vec{r}\vec{r}'}(t, t_0) + C^{(4B)}_{\vec{r}\vec{r}'}(t, t_0) - C^{(4C)}_{\vec{r}\vec{r}'}(t, t_0) - C^{(4D)}_{\vec{r}\vec{r}'}(t, t_0) \equiv 0 \,,$$
$$\text{if} \quad \vec{r} = 0 \text{ or } \vec{r}' = 0 \,. \tag{15}$$

In our simulation we choose the relative distance between the mesons to be the same at the initial and final times of the propagation, $\vec{r} = \vec{r}'$. In the spirit of the Born-Oppenheimer approximation it is assumed that the interaction proceeds faster than the motion between the mesons [11]. This seems to be justified because the quark mass $m_F = 0.1$ is relatively large. In this sense we replace the sum over all eigenstates in the spectral representation of the correlation function by an average term with an effective energy $W(\vec{r})$

$$C^{(4)}_{\vec{r}\vec{r}}(t, t_0) = \sum_n |\langle \vec{r}|n \rangle|^2 e^{-E_n(t-t_0)} \simeq c(\vec{r}) e^{-W(\vec{r})(t-t_0)} \,. \tag{16}$$

To the extent that the energy $W(\vec{r})$ of the meson-meson system at fixed relative distance $\vec{r}$ can be extracted from the large-$t$ behaviour of the diagonal elements we may conclude from (15)–(16) that $W(\vec{r} = 0) = +\infty$, provided the strength factor $c(\vec{r})$ is nonzero at $\vec{r} = 0$. In this case the effective interaction possesses a hard repulsive core. It is due to the anticommuting nature of the constituent fermion fields leading to Pauli repulsion.

6. *Results.* We want to test numerically the two ways suggested above through (13)–(14) and (16), respectively, to extract an effective meson-meson potential from a quantum field theory. The correlators $C^{(4)}$ and $\bar{C}^{(4)}$ were computed for time slices $t = 15 \ldots 19$ around the symmetry point $t_c = 17$ in periodic time. The potentials were obtained from cosh-fits to the correlators $\mathcal{C}$. The first definition (13)–(14) represents a matrix and leads to a nonlocal Hamilton operator $\mathcal{H}_I(\vec{r}, \vec{r}')$ where $\vec{r}, \vec{r}'$ is the separation between the mesons. Since computation of the entire 4-point correlation matrix is very time-consuming we only obtained the diagonal elements. This also allows a direct comparison with $\mathcal{W}_I(r) = W(r) - 2m$ from the second definition (16), where $2m = 1.031(9)$ is the mass of the noninteracting two-meson system.

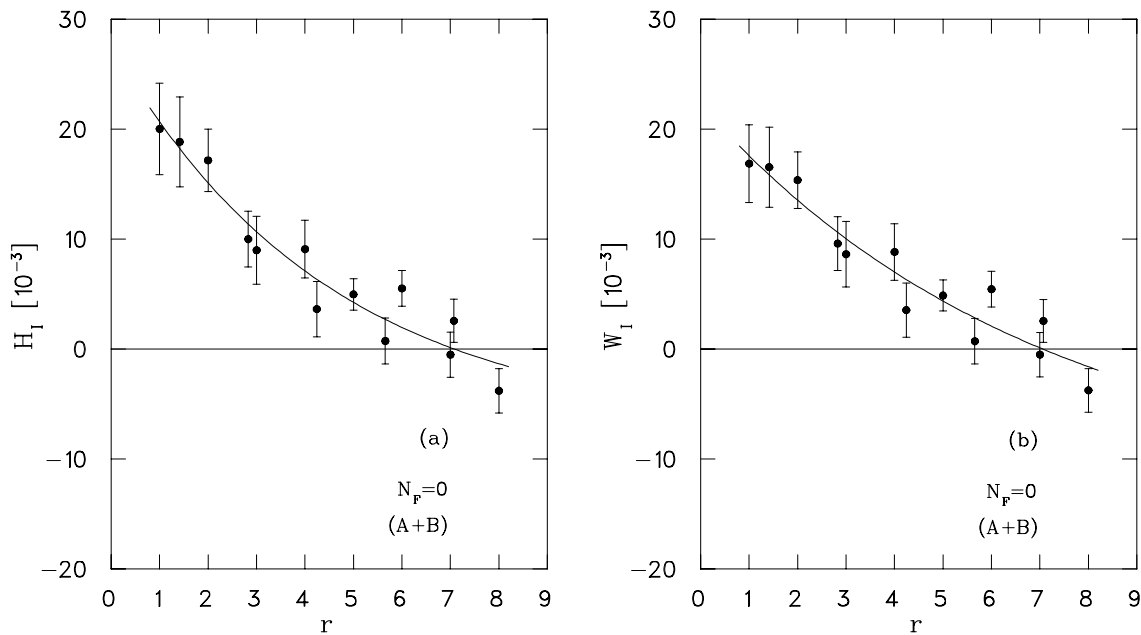

Figure 1: Meson-meson potentials as a function of relative distance $r$ for the direct channels (A+B) extracted from quenched $\mathrm{QED}_{2+1}$. The result (a) $\mathcal{H}_I(r)$ from the effective time evolution is compared with (b) $\mathcal{W}_I(r)$ of the adiabatic approximation. In both cases the direct channels lead to a repulsive potential barier. Curves are guide to the eye and error bars result from a jack-knife analysis.

We begin the description of our simulation data with the quenched theory, $N_F = 0$. The result for $\mathcal{H}_I(r)$ including the direct diagrams (A+B) only is presented in fig. 1a and that for $\mathcal{W}_I(r)$ is plotted in fig. 1b. Both results correspond to a repulsive potential barier. The quark-exchange diagrams (C+D) yield similar behaviour (not shown) except their contributions enter with a minus sign into the total correlations (6),(7).

In fig. 2 all four diagrams (A+B)–(C+D) are included. For both $\mathcal{H}_I(r)$ and $\mathcal{W}_I(r)$ there is some evidence for a repulsive core, as discussed earlier, followed by attraction at medium distances. The shapes emerge as a subtle interplay between all graphs (A+B)

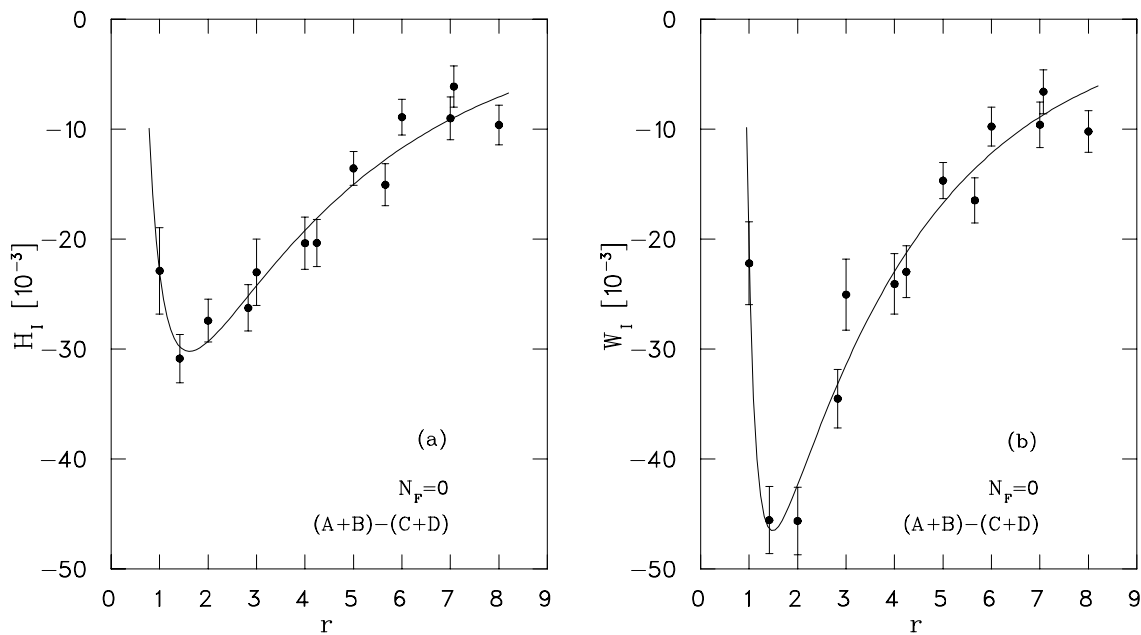

Figure 2: Full meson-meson potentials as a function of relative distance $r$ for all four graphs (A+B)–(C+D) in pure $\text{QED}_{2+1}$. Both total potentials (a) $\mathcal{H}_I(r)$ and (b) $\mathcal{W}_I(r)$ hint at a repulsive core at short range and exhibit attraction at medium distances. The resulting potential originates from a sensitive interplay between all contributing graphs. Curves are phenomenological fits to the data (see text) with jack-knife error bars.

and (C+D). We tried a phenomenological fit function of the type

$$\mathcal{V}_I(r) = \alpha r^{-\gamma} - \beta e^{-\delta r}, \tag{17}$$

and found the following parameters: $\alpha = 19.7[10^{-3}], \beta = 55.2[10^{-3}], \gamma = 2.35, \delta = 0.254$, for $\mathcal{H}_I$, and $\alpha = 37.2[10^{-3}], \beta = 81.4[10^{-3}], \gamma = 5.42, \delta = 0.316$, for $\mathcal{W}_I$. An exploratory solution of the radial Schrödinger equation in two dimensions to the corresponding potential form indicates a loosely bounded state. In both approximations additivity between the potential from the quark-exchange graphs (C+D) and that of the direct term (A+B) with respect to the total potential, $\mathcal{H}^{(A+B)} - \mathcal{H}^{(C+D)} \simeq \mathcal{H}^{(A+B)-(C+D)}$, is fulfilled reasonably well. This is a nontrivial point since the extraction of the potentials from the corresponding correlation functions is a nonlinear procedure.

Simulations with $N_F = 2$ external flavours of dynamical fermions also have been performed. In this case 16 random sources (instead of 32) were used to generate the quark propagators. Further, the potentials from the correlator ratios are smaller. For these reasons the data are considerably more noisy. Both simulations were performed at the same value of $\beta = 1.5$. For ease of comparison, in the following pictures we rescale the energy axis to that of the quenched computation by a factor equal to the mass ratio

$$\frac{2m(N_F = 0)}{2m(N_F = 2)} = \frac{1.031(9)}{0.63(1)} = 1.65 \,.$$ (18)

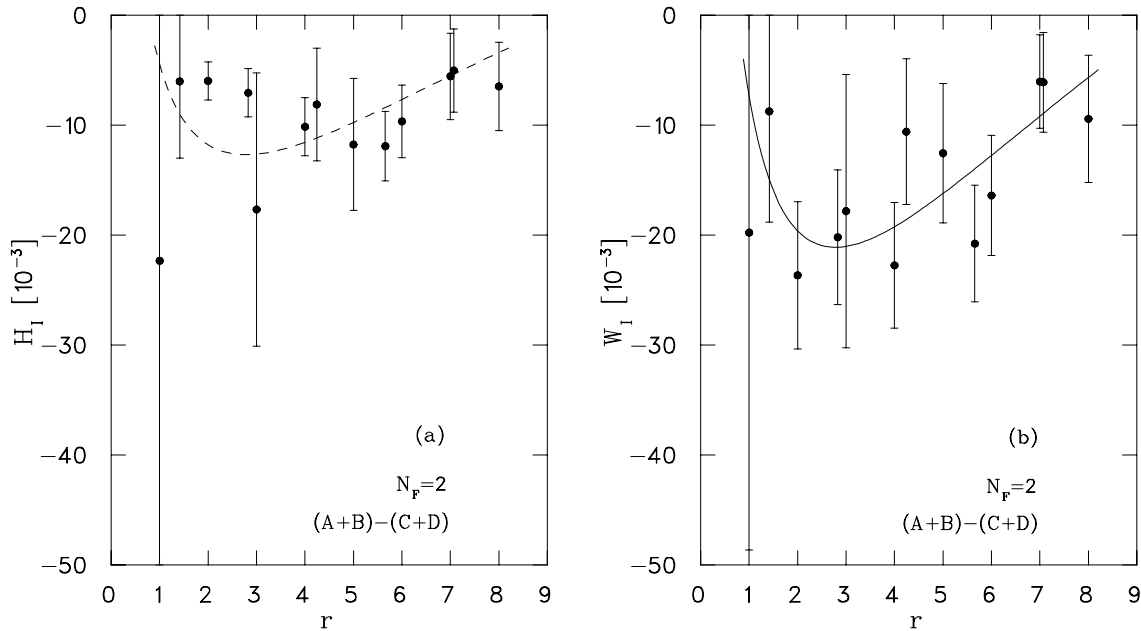

Figure 3: Full meson-meson potentials as a function of relative distance $r$ for all four graphs (A+B)–(C+D) extracted from $QED_{2+1}$ with two flavours. Both total potentials (a) $\mathcal{H}_I(r)$ and (b) $\mathcal{W}_I(r)$ resemble those of the pure gauge case. Dynamical fermions lead to softer interactions. The precision of the computation is not as accurate as in the quenched simulation.

In fig. 3a the results for the total potential $\mathcal{H}_I(r)$ are shown. There seems no qualitative change compared to the pure gluonic case, except that sea quarks apparently lead to a softer behaviour. The total potential $\mathcal{W}_I(r)$ from the adiabatic approximation is shown in fig. 3b. Again it resembles $\mathcal{H}_I$ and dynamical quarks lead to a certain softening of the interaction mechanism. Fits to the functional form (17) converged only for $\mathcal{W}_I$ (the dashed curve has simply been scaled).

*7. Conclusion.* We have proposed two possibilities to extract an effective interaction between two composite particles. This brings us closer to our aim of extracting poten-tials of two nucleons consisting of three quarks each from lattice QCD. Such a project necessitates the computation of a 4-point hadron Green function which is equivalent to a correlation function of six quark propagators within the QCD path-integral. A feasible approximation, which we study at present, might be to make two quarks of a baryon heavy so that they only act as spectator quarks in the scattering process [12].

This paper contains a first study of the influence of both gluons and dynamical quarks on meson-meson systems in the framework of $QED_{2+1}$. The sea quarks constitute the meson exchange in the sense of the Yukawa theory. With the extension of the pure gauge action to the full QED action the contribution of virtual mesons was investigated. It was found that gluon exchange and valence-quark exchange are equally important. Sea quark effects seem to be relatively small but might be decisive for the definite shape of the potential. In the model used mesons feel a short-ranged hard-core potential due the Pauli exclusion principle acting between their fermionic constituents. The medium-range residual forces between colour-neutral quark-clusters are attractive. It should be noted that a computation of meson-meson scattering phase shifts based on Lüscher's proposal [13] indicates the same behaviour of the potential [14]. The correct antisymmetrization of the meson-meson correlator was important for the emergence of a hard core. This is in contrast to earlier work [3] in QCD with *static* quarks where antisymmetrization is not possible. In the case of $SU(3)$ the nonabelian colour structure will not necessarily lead to a vanishing correlator, as for $U(1)$ in (15), when the relative nucleon-nucleon distance approaches zero. Preliminary studies seem to indicate attraction. Further investigations with light valence quarks are necessary to solve the hard-core problem of the nucleon-nucleon interaction.

*8. Acknowledgment.* We thank CEBAF where this work was initiated and finalized for the inspiring atmosphere. We are grateful to N. Isgur for valuable comments on the hard-core mechanism, and to H. Leeb for solving the Schrödinger equation for the parametrized potentials. Detailed discussions on the physical validity of the approxima-tions with W. Sakuler and C. Starkjohann are appreciated.